\documentclass{PoS}

\title{Why rooting fails}

\ShortTitle{Why rooting fails}

\author{\speaker{Michael Creutz}%
         \thanks{This manuscript has been
authored under contract number DE-AC02-98CH10886 with the
U.S.~Department of Energy.  Accordingly, the U.S. Government retains a
non-exclusive, royalty-free license to publish or reproduce the
published form of this contribution, or allow others to do so, for
U.S.~Government purposes.}\\
        Brookhaven National Laboratory\\
        E-mail: \email{creutz@bnl.gov}}

\abstract{I explore the origins of the unphysical predictions from
rooted staggered fermion algorithms.  Before rooting, the exact chiral
symmetry of staggered fermions is a flavored symmetry among the four
"tastes."  The rooting procedure averages over tastes of different
chiralities.  This averaging forbids the appearance of the correct 't
Hooft vertex for the target theory.}

\FullConference{The XXV International Symposium on Lattice Field
 Theory\\
		 July 30 - August 4 2007\\
		 Regensburg, Germany}

\begin{document}

\section{Introduction}

My presentation \cite{Creutz:2006wv} at last year's meeting in this
series discussed what were some of the unphysical consequences arising
from the rooting prescription usually used along with the staggered
quark formalism.  In particular, I showed how an excess symmetry leads
to an incorrect quark mass dependence.  That argument is elucidated
somewhat further in \cite{Creutz:2007yg,Creutz:2007nv}.  Here I
explore this problem more deeply to understand why these wrong
predictions appear.  In particular, I will demonstrate that a strong
mixing of tastes with different chiralities leads to an incorrect 't
Hooft vertex.

The outline is as follows.  In section \ref{infavor} I summarize the
naive arguments in favor of the rooting trick.  This includes treating
of the determinant as a sum over loops and empirical observations on
the eigenvalues of the Dirac operator on typical gauge configurations.
Section \ref{review} reviews the basic formulation of staggered
fermions and how the rooting is implemented.  Section \ref{awry} turns
to some issues related to chiral symmetry that signal caution.  In
particular, the method involves an averaging over fermion
chiralities. Also, on moving between topological sectors, the taste
symmetry of the eigenvalue spectrum must break.  In section
\ref{thooft} I connect these issues to an old topic, the `` 't Hooft
vertex.''  Here I show how symmetries forbid the rooting procedure
from correctly reproducing the requisite form.  Section
\ref{questions} raises a few questions, hinting at why the
approximation may not be too bad for the two light plus one
intermediate mass situation.  I also suggest some possible ways to
repair the algorithm.  Section \ref{conclusion} briefly states the
final conclusion that rooting can often be a good approximation but
predictions for non-perturbative physics where the 't Hooft vertex is
important can not be trusted.

\section{Naive justifications for rooting}\label{infavor}

The basic argument for rooting comes from consideration of the fermion
contribution as a determinant of the Dirac operator.  A determinant is
a sum over permutations of the rows of the matrix.  Each permutation
in turn is a product of cycles.  Each cycle represents a fermion loop
in perturbation theory.  Now with unrooted staggered fermions on a
smooth gauge field background, we have an excess in the number of
species by a factor of four.  Thus each loop is counted four times too
much.  By replacing the determinant with its fourth root, this
effectively multiplies each loop by a factor of one quarter, giving
exactly the desired contribution from a single fermion.  The
conclusion is that the rooting procedure does give the correct
perturbative expansion.  As a special case, rooting is correct for the
continuum limit of the free fermion theory as well.

This argument relies on symmetry between the four ``tastes'' of the
unrooted theory.  The importance of this is strongly emphasized in
Ref.~\cite{Bernard:2006vv}.  Numerical evidence for the required
symmetry appears in studies of the eigenvalues of the Dirac operator;
for example, Refs.~\cite{Durr:2004as} and \cite{Follana:2005km}, show
that as the lattice spacing becomes small the eigenvalues tend to
group into approximately degenerate quartets.  Rooting effectively
selects one eigenvalue from each quartet.
 
These arguments for rooting are further supported by the rather
spectacular successes of recent simulations.  Indeed, a variety of
observables that had previously been calculated in the ``valance
approximation'' have now been redetermined with dynamical quarks
treated using the fourth root approximation.  The agreement with
experiment is uniformly much better with the dynamical quarks
included.  For example, see Ref.~\cite{Davies:2003ik}.

Unfortunately these arguments have seduced many into suggesting that
the algorithm might become exact in the continuum limit, i.e. that the
lattice artifacts might go away as the lattice spacing is taken to
zero \cite{Sharpe:2006re}.  The purpose of this talk is to provide a
proof that this is impossible; there exist certain important physical
effects that the algorithm inherently must miss.

\section{Staggered review}\label{review}

To proceed I need to delve into the heart of the staggered algorithm.
For this I begin with the so called ``naive'' discretization of the
Dirac equation.  This considers fermions hopping between nearest
neighbor lattice sites while picking up a factor of $\pm i\gamma_\mu$
for a hop in direction $\pm \mu$.  Going to momentum space, the
discretization replaces powers of momentum with trigonometric
functions, for example
\begin{equation}
\gamma_\mu p_\mu\rightarrow \gamma_\mu {\sin(ap_\mu)\over a}.
\end{equation}
Here I denote the lattice spacing by $a$.  This formulation exposes
the famous ``doubling'' issue, arising because the fermion propagator
has poles not only for small momentum, but also whenever any component
of the momentum is at $\pi/a$.  The theory represents not one fermion,
but sixteen.

It is important to note that the various doublers have differing
chiral properties.  This arises from the simple observation that
\begin{equation}
{d\over dp}\sin(p)|_{p=\pi}=-{d\over dp}\sin(p)|_{p=0}.
\end{equation}
The consequence is that the helicity projectors $(1\pm\gamma_5)/2$ for
a travelling particle depend on which doubler one is observing.

\begin{figure*}
\centering
\includegraphics[width=1.6in]{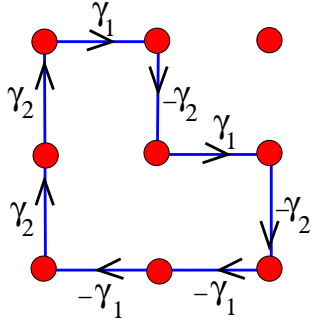}
\caption{When a fermion circumnavigates a loop in the naive
  formulation, it picks up a factor that always
  involves an even power of any particular gamma matrix.}
\label{loop} 
\end{figure*}

Now consider a fermion traversing a closed loop on the lattice.  As
shown in Fig.~\ref{loop}, the corresponding gamma matrix factors will
always involve an even number of any particular $\gamma_\mu$.  Thus
the resulting product is proportional to the identity.  If a fermion
starts in a single spinor component, it will wind up in the same
component after circumnavigating the loop.  The determinant exactly
factorizes into four equivalent pieces.  The naive theory has an exact
$U(4)$ symmetry, as pointed out some time ago by Karsten and Smit
\cite{Karsten:1980wd}.  Indeed, for massless fermions this is actually
a $U(4)\otimes U(4)$ chiral symmetry.  This symmetry does not
contradict any anomalies since it is not the full naive $U(16)\otimes
U(16)$ of 16 species.  The chiral symmetry generated by $\gamma_5$
remains exact, but this is allowed because it is actually a flavored
chiral symmetry. As mentioned above the helicity projectors for the
various doubler species use different signs for $\gamma_5$.

The basic idea of staggered fermions is to divide out this $U(4)$
symmetry \cite{Kogut:1974ag,Susskind:1976jm,Sharatchandra:1981si}  by
projecting out a single component of the fermion spinor on each site.
Taking $\psi\rightarrow P\psi$, one projector that accomplishes this
is
\begin{equation}
P={1\over 4} \bigg(1+i\gamma_1\gamma_2 (-1)^{x_1+x_2}
+i\gamma_3\gamma_4 (-1)^{x_3+x_4}
 +\gamma_5(-1)^{x_1+x_2+x_3+x_4}\bigg)
\label{projection}
\end{equation}
where the $x_i$ are the integer coordinates of the respective lattice
sites.  This immediately reduces the doubling from a factor of sixteen
to four.

At this stage the naive $U(1)$ axial symmetry remains.  Indeed, the
projector used above commutes with $\gamma_5$.  This symmetry is
allowed since four species, usually called ``tastes,'' remain.  Among
them the symmetry is a taste non-singlet; under a chiral rotation, two
rotate one way and two the other.

The next step taken by most of the groups using staggered fermions is
the rooting trick.  In the hope of reducing the multiplicity down from
four, the determinant is replaced with its fourth root,
$|D|\rightarrow |D|^{1/4}$.  With several physical flavors this trick
is applied separately to each.  As argued in the previous section, in
simple perturbation theory each fermion loop gets multiplied by one
quarter, cancelling the extra factor from the four ``tastes.''

At this point one should be extremely uneasy: the exact chiral
symmetry is waving a huge red flag.  Symmetries of the determinant
survive rooting, and thus the exact $U(1)$ axial symmetry for the
massless theory remains.  For the unrooted theory this was a flavored
chiral symmetry.  But, having reduced the theory to one flavor, how
can there be a flavored symmetry without multiple flavors?

Here I need to make a somewhat technical comment on chiral symmetry.
It is usually regarded in terms of an $SU(N_f)\otimes SU(N_f)$
symmetry for the massless $N_f$ flavor theory.  This is believed to be
spontaneously broken, and, via the Goldstone mechanism, explains the
lightness of the pions.  However, this is also a symmetry of the {\bf
massive} theory regarded in terms of its parameter space.  More
specifically, consider a mass term of form
\begin{equation}
{1\over 2}(\overline\psi_L M \psi_R+\overline\psi_R M^\dagger\psi_L)
\end{equation}
where $M$ is a complex $N_f$ by $N_f$ matrix.  Then physics is
invariant under changing the mass parameters
\begin{equation}
M\rightarrow g_L^\dagger M g_R.
\end{equation}
Here $g_L$ and $g_R$ are arbitrary matrices in $SU(N_f)$.

In this context it is important to note that the theory is {\bf not}
invariant under a simple phase change $M\rightarrow e^{i\theta}M$.
Such a rotation is anomalous and changes the strong $CP$ violating
angle.  This is the reason there are only $N_f^2-1$, rather than
$N_f^2$, Goldstone bosons.  In the particular case of one-flavor QCD,
there should be no surviving chiral symmetry whatever.  That theory is
expected to be analytic in the fermion mass in the vicinity of the
origin \cite{Creutz:2006ts}.  Unfortunately, a phase change in the
mass term is an exact symmetry of the staggered fermion determinant
and remains so on rooting.  This incorrect behavior was the main
subject of my discussion last year \cite{Creutz:2006wv}.

\section{Where things go awry}\label{awry}

So the rooted theory appears to have some issues related to chiral
symmetry and the anomaly.  Before rooting, the one exact chiral
symmetry is actually a non-singlet symmetry because the different
tastes are associated with different gamma matrix conventions.  Thus
there are two tastes of each chirality.  What happens to this symmetry
on rooting?

Consider the index theorem for a gauge configuration with unit
winding.  Near the continuum limit there should be one approximate
zero mode for each of the tastes.  These are not exact zero modes
because of finite spacing effects, but that is not the issue here.
Because the tastes differ in chirality, two of these modes will be
left handed and two right handed in the sense of the physical helicity
projectors for the corresponding fermions.  What rooting does is
average over these.  While this allows the chiral symmetry to remain,
it does not correspond to the single chirality mode of the target
theory.  The issue is analogous to trying to make a living organism
out of a racemic mixture of proteins; it won't work.

\begin{figure*}
\centering
\includegraphics[width=2in]{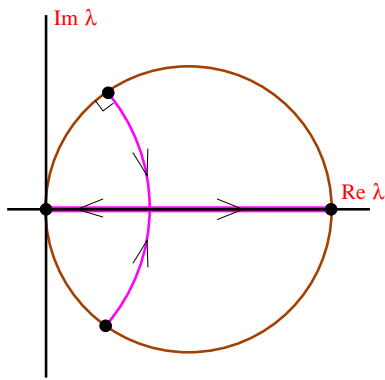}
\caption{In the overlap formulation a single exact zero eigenmode is
  possible.  In transiting from winding number zero to one, a pair of
  complex eigenvalues disappear and are replaced with the exact zero
  mode and a compensating mode on the opposite side of the overlap circle.}
\label{overlap} 
\end{figure*}

Note that the staggered projection operator in Eq.~\ref{projection}
satisfies
\begin{equation}
P\gamma_5=\gamma_5 P=(-1)^{x_1+x_2+x_3+x_4}P.
\end{equation}
This means that the oscillating factor $(-1)^{x_1+x_2+x_3+x_4}$ plays
the role of $\gamma_5$.  This matrix is independent of gauge
configuration and as such remains traceless independent of the gauge
field winding number.  This is another way to see that the approximate
zero modes must come in opposite chiralities.

This behavior is unlike that with other formulations.  In usual
``continuum'' discussions, the appearance of zero modes is compensated
by modes that move in from infinity.  With Wilson fermions the chiral
zero modes are paired with heavy doubler states.  With the overlap
operator \cite{Neuberger:1997fp}, a zero mode has a compensating mode
occurring on the opposite side of the overlap circle.  This behavior
with the overlap is sketched in Fig.~\ref{overlap}, taken from my
Lattice '02 presentation \cite{Creutz:2002qa}.  The overlap
formulation introduces a modified chirality matrix $\hat\gamma_5$
which does depend on the gauge fields.  The winding number is given by
the relation $\nu={\rm Tr}\ \hat\gamma_5/2.$

\begin{figure*}
\centering
\includegraphics[width=2.5in]{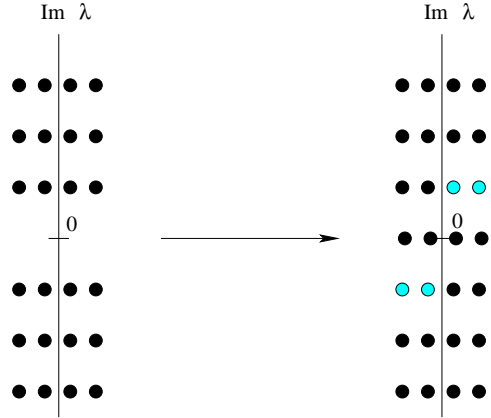}
\caption{In transiting between different winding number sectors, the
  clustering of the staggered Dirac eigenvalues into taste quartets
  must break down.}
\label{transiting} 
\end{figure*}

The approximate zero modes have implications for how the staggered
eigenvalues must evolve as one moves between topological sectors.  For
a smooth gauge field with zero winding number, near the continuum
limit there is numerical evidence that the Dirac eigenvalues indeed
cluster into taste quartets.  A similar structure is desired with
smooth gauge fields carrying a unit of winding, although in this case
there should be one quartet of approximate zero modes.  However on
considering rough gauge fields that interpolate between these
situations, the quartets must necessarily break apart.  In particular,
two of the approximate zero modes must drop down from above in the
complex eigenvalue plane, while two rise up from below.  This
necessarily leaves a mismatch in the form of ``holes'' that must be
absorbed in the non-zero eigenvalue spectrum, as sketched in
Fig.~\ref{transiting}.

Note that, despite claims to the contrary \cite{Bernard:2006vv}, this
chiral mixing has nothing to do with the order of taking the continuum
limit and going to zero mass.  Even when the mass remains finite, a
topologically non-trivial gauge configuration should still generate
fermion eigenvalues with approximately zero imaginary part.  The mass
merely gives these modes a finite real part.  I do, of course, assume
that the lattice spacing is small enough that the chiral modes
corresponding to topology are clearly identifiable.

\section{The 't Hooft vertex}\label{thooft}

So the concerns with rooting involve zero modes of the massless Dirac
operator.  Through the index theorem, this in turn is tied to the
topological structure of the gauge field.  To explore the consequences
of this connection, start with the usual integration of the fermionic
fields in terms of determinant of the Dirac operator, $D$.  For any
given configuration of gauge fields, this determinant is the product
of the eigenvalues of this matrix.  To control infrared issues, insert
a small mass and write the resulting path integral
\begin{equation}
Z=\int dA\ 
e^{-S_g}\ 
\prod_i (\lambda_i+m). 
\end{equation}
Here the $\lambda_i$ are the eigenvalues of the kinetic part of the
fermion determinant and $S_g$ is the pure gauge part of the action.
On taking the mass to zero, any configurations which contain a zero
eigenmode will have zero weight in the path integral.  This suggests
that for the massless theory one can ignore any instanton effects
since the corresponding configurations don't contribute to the path
integral.  Does this mean that ``instantons'' are irrelevant in the
continuum limit?

Indeed, 't Hooft \cite{'tHooft:1976up,'tHooft:fv} pointed out long ago
why this conclusion is incorrect.  The issue is not whether the zero
modes contribute to the path integral, but whether they can contribute
to physical correlation functions.  To see how this goes, add some
sources to the path integral
\begin{equation}
Z(\eta,\overline\eta)=\int dA\ d\psi\ d\overline\psi\  
e^{-S_g+\overline\psi (D+m) \psi +\overline\psi \eta+ \overline\eta\psi}.
\end{equation}
Differentiation (in the Grassmann sense) with respect to $\eta$ and
$\overline \eta$ gives any desired fermionic correlation function.
Now integrate out the fermions
\begin{equation}
Z=\int dA\ 
e^{-S_g-\overline\eta(D+m)^{-1}\eta}\ 
\prod_i (\lambda_i+m).
\end{equation}
Consider a source that overlaps with an eigenvector of $D$
corresponding to one of the zero modes, i.e.
\begin{equation}
(\psi_0,\eta)\ne 0.
\end{equation}
The source contribution introduces a $1/m$ factor into the path
integral.  This cancels the $m$ from the determinant, leaving a finite
contribution as $m$ goes to zero.

With multiple flavors, the determinant will have a mass factor from
each.  When several masses are taken to zero together, one will need a
similar factor from the sources for each.  This product of source
terms is the famous ```t Hooft vertex.''
\cite{'tHooft:1976up,'tHooft:fv} While it is correct that instantons
do drop out of $Z$, they survive in correlation functions.

\begin{figure*}
\centering
\includegraphics[width=2.5in]{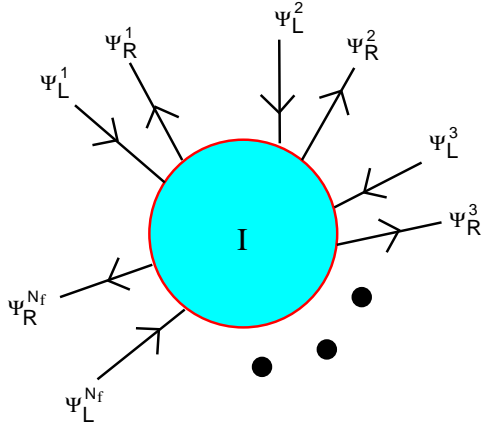}
\caption{With $N_f$ flavors the 't Hooft vertex is a $2N_f$ fermion
  interaction where each flavor flips its spin.}
\label{instanton} 
\end{figure*}

So, with $N_f$ flavors the theory generates a $2N_f$-fermion effective
interaction, as sketched in Fig.~\ref{instanton}.  This is a purely
non-perturbative phenomenon. In this interaction, all flavors flip
their spin, and this forms the basis of the anomaly.  With several
flavors this is a high dimensional operator, but it remains relevant
since the high dimensions are compensated by powers of the strong
interaction scale, $\Lambda_{qcd}$.  Indeed, the resulting interaction
is non-local at this scale.

For the unrooted staggered theory with its four tastes, the expected
't Hooft vertex is an octilinear interaction $\sim
(\overline\psi\psi)^4$.  It strongly couples all the tastes, even in
the continuum limit.  It does not violate the exact chiral symmetry of
the theory because it involves two tastes of each chirality.

Now for the target one-flavor theory, the 't Hooft vertex should
reduce to a simple bilinear interaction $\sim \overline\psi\psi$.
This has the form of a mass shift and is inconsistent with any exact
chiral symmetry, including that of the unrooted theory.  Indeed, that
symmetry forbids the generation of such a term in the rooted theory.
The absence of this vertex in the rooted approximation has serious
consequences; in particular, the rooted theory will have an incorrect
renormalization group flow for the fermion mass.

One might try to argue that since the unrooted determinant goes as the
mass to the fourth power, the rooted formula goes only linearly in the
mass.  Then to cancel the zero only requires one taste source.  So why
not just measure the vertex for one taste and ignore the others?

Unfortunately this will not work.  The basic vertex strongly couples
all tastes.  In the unrooted theory the strength of this coupling
scales as the product of $m^{-4}$ from the sources times $m^4$ from
the determinant.  This leaves a mass independent contribution.
However the rooted theory still has the $m^{-4}$ factor from the
sources but only $(m^4)^{1/4}\sim m$ from the rooted determinant.
Thus the rooted vertex displays a $m^{-3}$ singularity at vanishing
mass.  The scale of this singularity is set by $\Lambda_{qcd}$ and
there is no lattice spacing suppression.  Note also that high gluon
momenta are not involved, unlike the taste mixing arising in
perturbation theory.

\begin{figure*}
\centering
\includegraphics[width=3.5in]{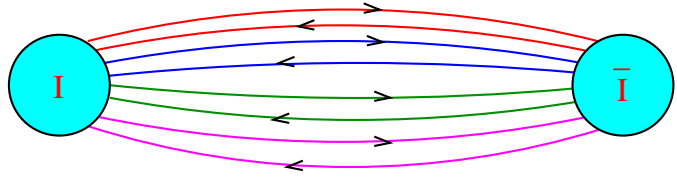}
\caption{In calculating the instanton/anti-instanton interaction, a
  contribution will arise from the exchange of all tastes.  This
  introduces an unphysical singularity in the rooted theory.}
\label{iibar} 
\end{figure*}

Because of this strong coupling between the tastes, all four must be
considered in intermediate states.  This will give unphysical
contributions to quantities such as multi-instanton interactions.  To
be more explicit, imagine trying to calculate a correlation function
of two pseudo-scalar glue-ball operators $F\tilde F(x)$ and $ F\tilde
F(y)$.  This will receive a contribution from instanton/anti-instanton
pairs.  That, in turn, will have a contribution from the exchange of
all four tastes, as sketched in Fig.~\ref{iibar}.  In the unrooted
theory the contribution of this diagram scales as the mass to the
zeroth power and has a spatial dependence arising from the overlap of
the approximate fermion zero modes of the two topological objects.  On
rooting the determinant factors associated with the instantons have a
reduced mass dependence, leaving behind a $m^{-6}$ mass dependence for
this correlation.  This is dramatically different from the desired
target one-flavor theory, where the exchange of the single physical
fermion scales again as a mass independent constant.  Note that for
the physical case the exchange of four copies of the fermion in the
zero mode is forbidden by the Pauli principle.

\section{Questions}\label{questions}

Several open questions remain on the rooting procedure.  Would a
square root of the determinant be better?  In particular, the doublers
do occur in equivalent pairs.  Also the reduction from four to two
flavors should leave behind a residual chiral symmetry; so, the exact
symmetry is not in itself necessarily bad.  The detailed form of the
't Hooft vertex will still couple the extra tastes, but this might be
a small effect.

Are these issues all associated with light quarks and could the wrong
chiral behavior of the rooting become unimportant for massive quarks?
Indeed, if this is the case and the square root is also a better
approximation, then the numerical successes of staggered fermions for
the two light plus one intermediate-mass-quark case would be easier to
understand.

Can counter-terms fix things?  One might try to add a counter-term to
mimic the desired 't Hooft vertex and another to cancel the unphysical
singularity.  This would require some tuning of the strengths of the
counter-terms. Also, given the non-local nature of the 't Hooft
vertex, it is unclear whether these terms would have to be non-local.
But perhaps this would be an acceptable price to pay for the gained
efficiency of the staggered approach.

Instead of rooting, it might be possible to cancel the extra tastes
with bosonic ghosts.  To avoid the unphysical averaging over
chiralities, this would require a chiral formulation for the ghosts.
But, since they are bosonic, this might not impose the computational
costs of directly simulating chiral fermions.

\section{Conclusion}\label{conclusion}

The conclusions of this discussion are quite succinctly stated.
First, rooting is a justified perturbative procedure.  As such, it can
be accurate for many physical quantities.  However it cannot become
exact in the continuum limit because it does not generate the correct
't Hooft vertex.  This makes the scheme particularly dangerous for the
treatment of non-perturbative physics in singlet channels.


\begin{thebibliography}{99}

\bibitem{Creutz:2006wv}
  M.~Creutz,
  arXiv:hep-lat/0608020.

\bibitem{Creutz:2007yg}
  M.~Creutz,
  Phys.\ Lett.\  B {\bf 649} (2007) 230
  [arXiv:hep-lat/0701018].

\bibitem{Creutz:2007nv}
  M.~Creutz,
  Phys.\ Lett.\  B {\bf 649} (2007) 241
  [arXiv:0704.2016].

\bibitem{Bernard:2006vv}
  C.~Bernard, M.~Golterman, Y.~Shamir and S.~R.~Sharpe,
  Phys.\ Lett.\  B {\bf 649} (2007) 235
  arXiv:hep-lat/0603027.

\bibitem{Durr:2004as}
  S.~Durr, C.~Hoelbling and U.~Wenger,
  Phys.\ Rev.\  D {\bf 70} (2004) 094502
  [arXiv:hep-lat/0406027].

\bibitem{Follana:2005km}
  E.~Follana, A.~Hart, C.~T.~H.~Davies and Q.~Mason  [HPQCD Collaboration],
  Phys.\ Rev.\  D {\bf 72} (2005) 054501
  [arXiv:hep-lat/0507011].

\bibitem{Davies:2003ik}
  C.~T.~H.~Davies {\it et al.}  [HPQCD Collaboration],
  Phys.\ Rev.\ Lett.\  {\bf 92} (2004) 022001
  [arXiv:hep-lat/0304004].

\bibitem{Sharpe:2006re}
  S.~R.~Sharpe,
  PoS {\bf LAT2006} (2006) 022
  [arXiv:hep-lat/0610094].

\bibitem{Karsten:1980wd}
  L.~H.~Karsten and J.~Smit,
  Nucl.\ Phys.\  B {\bf 183} (1981) 103.

\bibitem{Kogut:1974ag}
  J.~B.~Kogut and L.~Susskind,
  Phys.\ Rev.\ D {\bf 11}, 395 (1975).

\bibitem{Susskind:1976jm}
  L.~Susskind,
  Phys.\ Rev.\ D {\bf 16}, 3031 (1977).

\bibitem{Sharatchandra:1981si}
  H.~S.~Sharatchandra, H.~J.~Thun and P.~Weisz,
  Nucl.\ Phys.\ B {\bf 192}, 205 (1981).

\bibitem{Creutz:2006ts}
  M.~Creutz,
  Annals Phys. {\bf 322} 1518 (2007)
  [arXiv:hep-th/0609187].

\bibitem{Neuberger:1997fp}
H.~Neuberger,
Phys.\ Lett.\ B {\bf 417}, 141 (1998)
[arXiv:hep-lat/9707022].

\bibitem{Creutz:2002qa}
  M.~Creutz,
  Nucl.\ Phys.\ Proc.\ Suppl.\  {\bf 119} (2003) 837
  [arXiv:hep-lat/0208026].

\bibitem{'tHooft:1976up}
  G.~'t Hooft,
  Phys.\ Rev.\ Lett.\  {\bf 37} (1976) 8.

\bibitem{'tHooft:fv}
G.~'t Hooft,
''
Phys.\ Rev.\ D {\bf 14}, 3432 (1976)
[Erratum-ibid.\ D {\bf 18}, 2199 (1978)].

\end{thebibliography}
\end{document}